\documentclass[a4paper,11pt]{article}
\usepackage{pos}
\newcommand{\qmc}[1]{QMC$\pi$-#1}

\title{Predictions in the superheavy region from the\\quark-meson coupling model \qmc{III}}

\author*[a]{Kay Marie M. Paglinawan}
\author[b]{Anthony W. Thomas}
\author[b]{Pierre A. M. Guichon}
\author[c]{Jirina R. Stone}

\affiliation[a]{Physics Department, Silliman University, Hibbard Avenue, Dumaguete City, Philippines}

\affiliation[b]{CSSM and ARC Centre of Excellence for Dark Matter Particle Physics,\\
Department of Physics, University of Adelaide, SA 5005, Australia}

\affiliation[c]{Department of Physics (Astrophysics), University of Oxford, Oxford OX1 3RH, UK}

\emailAdd{kaylmartinez@su.edu.ph}
\emailAdd{anthony.thomas@adelaide.edu.au}

\abstract{The Quark-Meson Coupling (QMC) model establishes a self-consistent relationship between the quark structure dynamics of a nucleon and the relativistic mean fields that arise within the nuclear medium \cite{Guichon2018}. The model has been successful in calculating the ground-state observables of finite nuclei and in predicting the properties of dense nuclear matter, as well as cold, nonaccreting neutron stars \cite{Stone2016, Stone2017, Martinez2019, Martinez2020, Martinez2020e, Thomas2023, Antic2020}. This paper focuses on the latest predictions from the model for the superheavy region, encompassing energies and deformations. Despite utilizing significantly fewer model parameters, the results have consistently improved as the model evolved, yielding better predictions for binding energies.}

\FullConference{The XVIth Quark Confinement and the Hadron Spectrum Conference (QCHSC24)\\
 19-24 August, 2024\\
 Cairns Convention Centre, Cairns, Queensland, Australia\\}

\begin{document}
\maketitle

\section{Introduction}
Exploration beyond the present boundaries of the nuclear landscape remains an ongoing challenge in both theoretical and experimental atomic and nuclear physics. Several theoretical nuclear models have been developed to reproduce various nuclear observables and to make predictions for the uncharted regions of the nuclear chart. In the region of superheavy elements (SHEs), numerous laboratories worldwide are dedicated to synthesizing new elements and isotopes despite the arduous "atom-at-a-time" discovery process \cite{Smits2024}. In particular, searches for new elements beyond Oganesson (with a proton number $Z=118$) and near the theorized next neutron magic number $N=184$ are exciting and dynamic research areas. Although these investigations are quite experimentally challenging, they are crucial for establishing the existence or non-existence of the long-speculated "island of stability" \cite{Nazarewicz2018, Smits2024}.

The Quark-Meson Coupling (QMC) model has successfully reproduced experimental data on the ground-state properties of even-even finite nuclei across the nuclear chart \cite{Martinez2020, Martinez2019, Martinez2020e, Thomas2023}. Throughout its development, the model has consistently improved predictions for energies, nuclear sizes, and shapes, successfully capturing salient features of finite nuclei as seen in available experimental data. The earlier version of the model, \qmc{I}, was used to predict binding energies and deformations for SHEs\cite{Stone2019}. In this paper, we revisit the SHE region using the latest version of the model, \qmc{III}, which features a reduced number of parameters and incorporates additional physical contributions.

\section{Theory}\label{theory}
The latest quark-meson coupling model, \qmc{III}, is discussed in \cite{Martinez2020}. In this section, we review the relevant theory of the QMC framework and its developments used in the computations presented in this paper.

In the QMC model, nucleons are constructed at the quark level, and their interactions are described through meson exchange. The MIT bag model is employed to characterize each nucleon as containing three quarks, with the nucleon bags considered to be non-overlapping. As a result of these interactions, the baryon mass is modified into an effective mass through the equation:
\begin{equation}
M^{\text{*}}_B = M_B-g_\sigma\sigma + \frac{d}{2}( g_\sigma\sigma)^2.
\end{equation}
Here, $M^*_B$ and $M_B$ are the effective and original baryon masses, respectively. $\sigma$ denotes the $\sigma$ meson field, $g_\sigma$ represents the quark coupling to the $\sigma$ meson,  and $d$ is the scalar polarisability unique to the QMC model. In addition to the \(\sigma\) meson, which accounts for intermediate-range attraction, the model also includes the vector meson \(\omega\) to represent short-range repulsion and the vector-isovector \(\rho\) meson to account for isospin dependence.

The classical total energy is expressed in terms of the energies of the quarks and mesons, leading to the derivation of the QMC Hamiltonian by eliminating the meson fields. Following this, the QMC energy density functional (EDF) is derived using Hartree-Fock calculations.

The QMC EDF contains five adjustable and well-constrained parameters:
$G_{\mu_q} = \frac{g_\mu}{m_\mu^2}$ ($\mu = \sigma, \omega, \rho$),
$m_\sigma$, and $\lambda_3$, where $g_\mu$ are the couplings of the quarks to the mesons, $m_\mu$ the meson masses and  $\lambda_3$ accounts for the self-coupling of the $\sigma$ meson.

\subsection{QMC-derived pairing functional}\label{pairing}
In the earlier versions of the QMC model, nuclear pairing was implemented in a standard manner using Hartree-Fock + Bardeen-Cooper-Schrieffer (HF+BCS) volume pairing, with two additional pairing strength parameters optimized alongside the QMC parameters. A significant advancement in the latest version of the QMC model is the implementation of pairing based on Bogoliubov theory, allowing the pairing functional to be fully expressed in terms of the existing QMC parameters. This change eliminates the need for the two extra parameters. Furthermore, the QMC pairing functional is now naturally density-dependent, in contrast to the volume pairing used in earlier versions, which does not account for density variations.

\subsection{Other contributions to the total EDF}\label{other}
In addition to the mean field QMC functional, the total EDF includes other contributions. These are the single-pion exchange, which is evaluated using the local density approximation, and the Coulomb interaction, which is expressed in a standard form that includes both direct and exchange terms. Additionally, a center-of-mass correction is applied. These functionals are consistent with those used in \qmc{II}, and for more details, the reader is referred to Ref. \cite{Martinez2019}.

\section{Method}\label{method}
The QMC EDF is solved using an axially symmetric HF+BCS code adapted by P.-G. Reinhard~\cite{skyax, Stone2016} to solve the finite nuclei. The model parameters are optimized using POUNDerS~\cite{Kort2010}, a derivative-free optimization procedure. The parameter search involves fitting to the available binding energies ($BE$) and root-mean-square charge radii (\(R_{ch}\)) of seventy magic and semi-magic nuclei, selected as described in \cite{Martinez2019}. In this updated version of the model, a total of 129 data points are used to fit the five parameters of \qmc{III} discussed in Section \ref{theory}. Once the optimised parameters were determined, calculations were performed for the ground-state properties of several nuclei in the superheavy region with \(Z \ge 100\). To evaluate the predictions against available data, residuals are computed by taking the difference between QMC results and that of experiment using the latest Atomic Mass Evaluation (AME) 2020 \cite{AME2020}.

\section{Results and discussion}
This section presents the results from calculations using \qmc{III} in the SHE region for isotopes with even number of protons and even number of neutrons (even $Z$, even $N$). In this latest version of the model, the parameter values are $G_\sigma = 9.62$ fm$^2$, $G_\omega=5.21$ fm$^2$, $G_\rho=4.71$ fm$^2$, $M_\sigma=503$ MeV, and $\lambda_3=0.05$ fm$^{-1}$. It is once again emphasized that there are only five parameters, as the pairing strengths are already fully expressed in terms of the existing QMC parameters.

The nuclear matter properties (NMPs) corresponding to the optimized \qmc{III} parameters fall within the acceptable range of values. The NMP values are as follows: saturation density $\rho_0 = 0.15$ fm$^{-3}$, saturation energy $E_0 = -15.7$ MeV, symmetry energy $S_0=29$ MeV, slope of symmetry energy $L_0 = 43$ MeV, and incompressibility of nuclear matter $K_0 = 233$ MeV. It is noteworthy that in the most recent two versions of the QMC model, the value of \( K_0 \) has significantly decreased, bringing it into the acceptable range compared to earlier versions. Additionally, the value of \( L_0 \) has increased in the newer QMC versions, aligning it more closely with data from terrestrial and astrophysical observations in Ref. \cite{Li2013}.

\subsection{Binding energies}

\begin{figure}[tbh!]
		\centering
	\begin{tabular}{c c}
		\includegraphics[angle=0,width=0.485\textwidth]{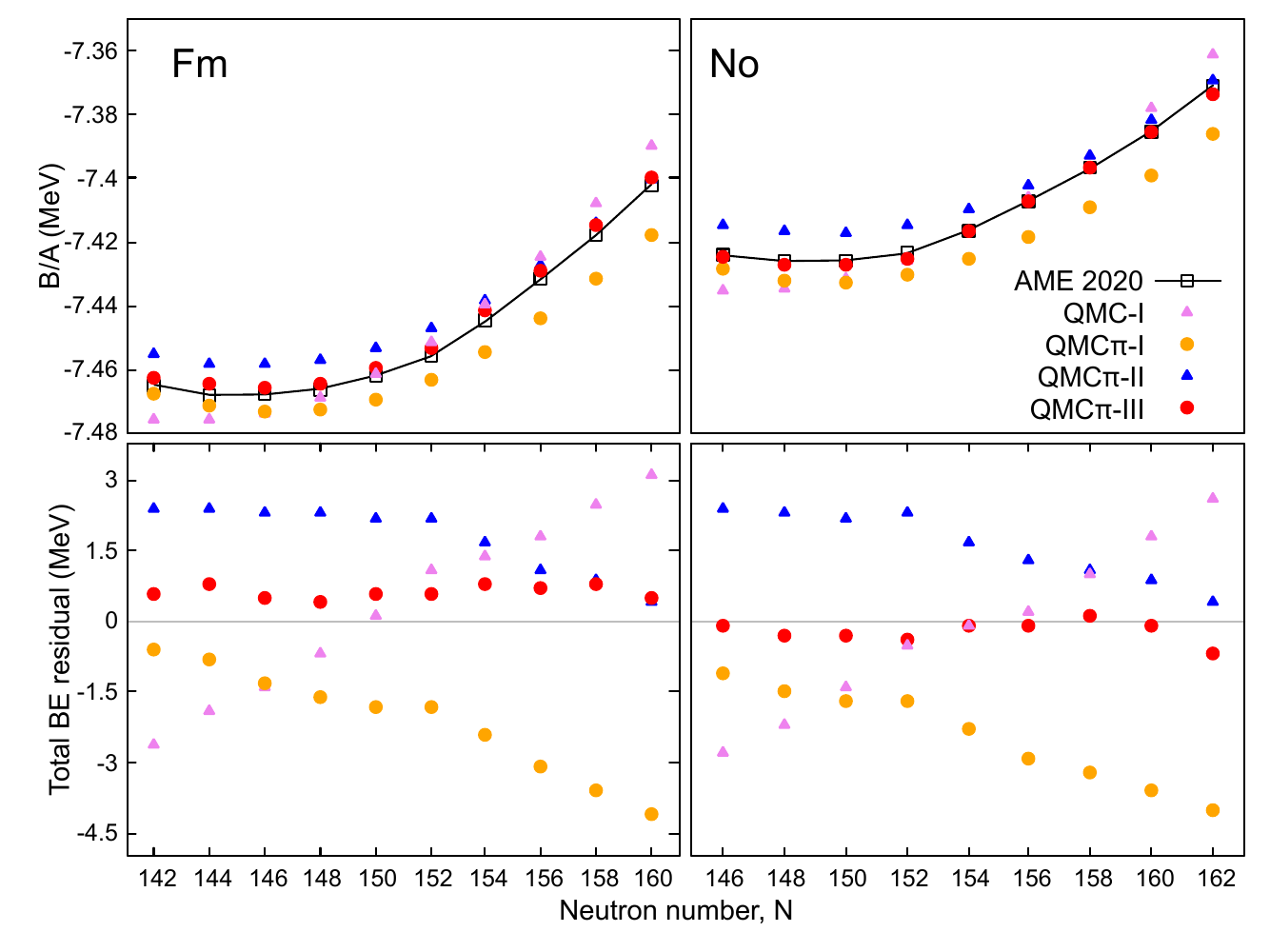}&
		\includegraphics[angle=0,width=0.485\textwidth]{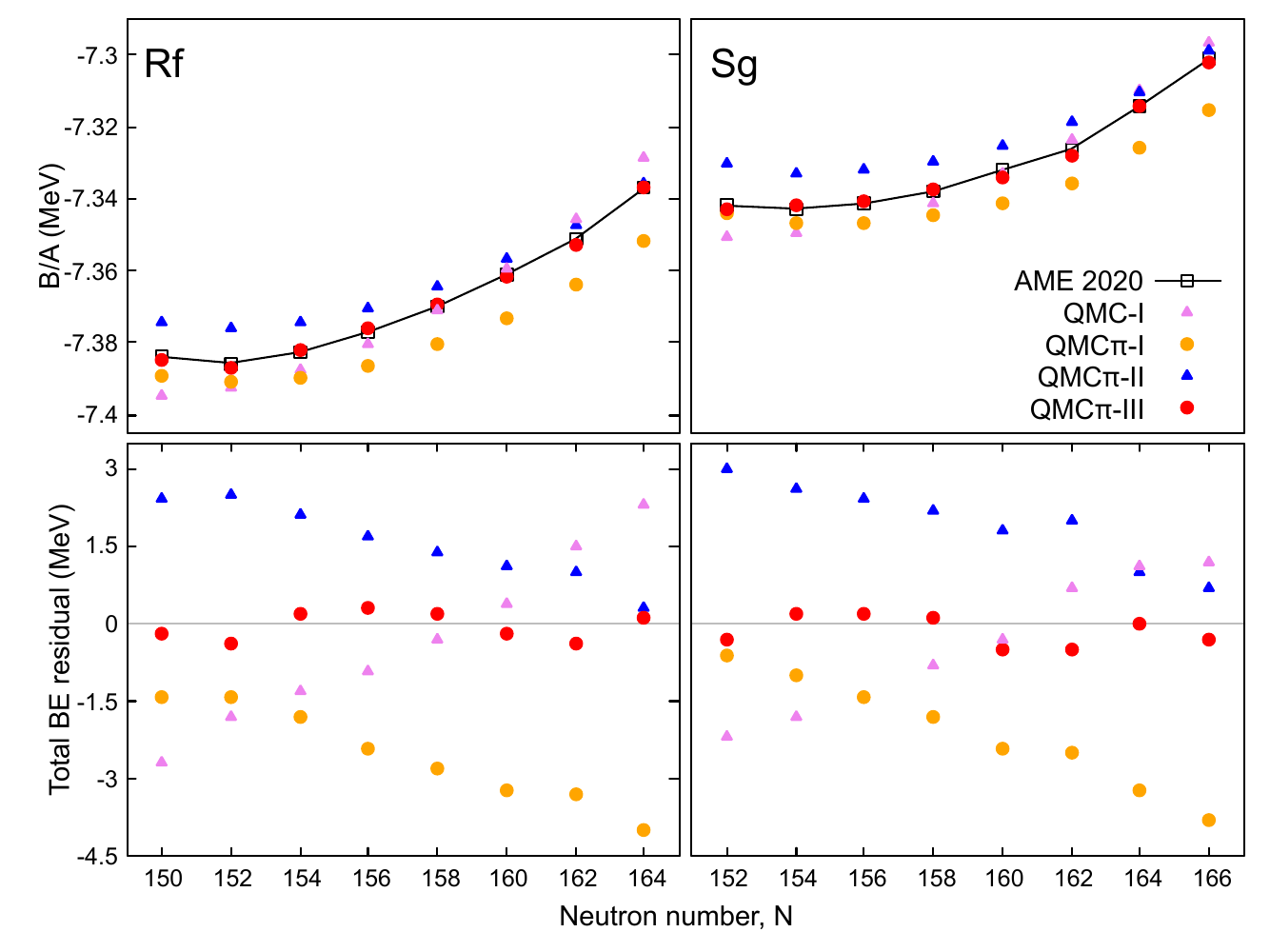}\\
		\includegraphics[angle=0,width=0.485\textwidth]{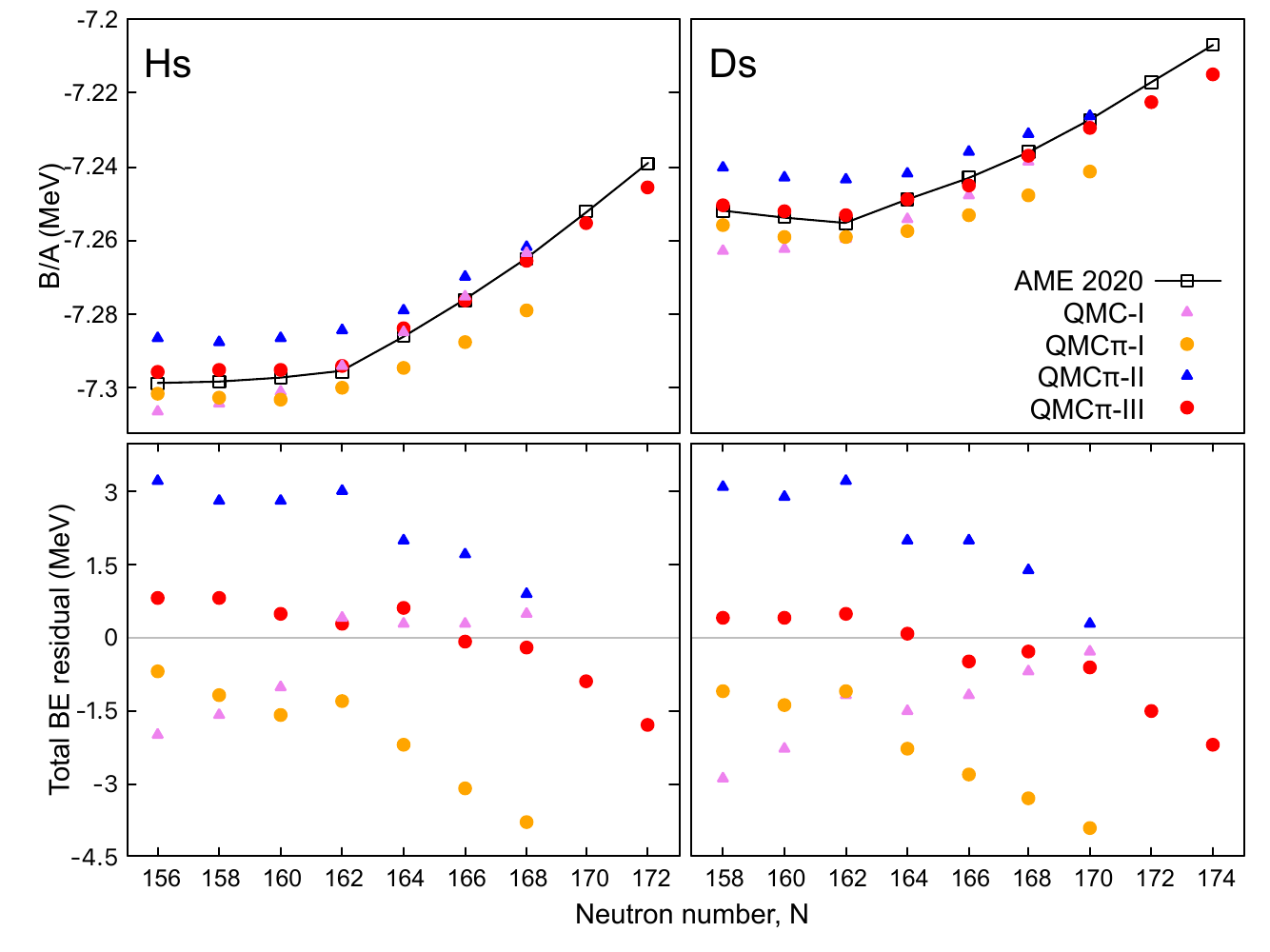}&
		\includegraphics[angle=0,width=0.485\textwidth]{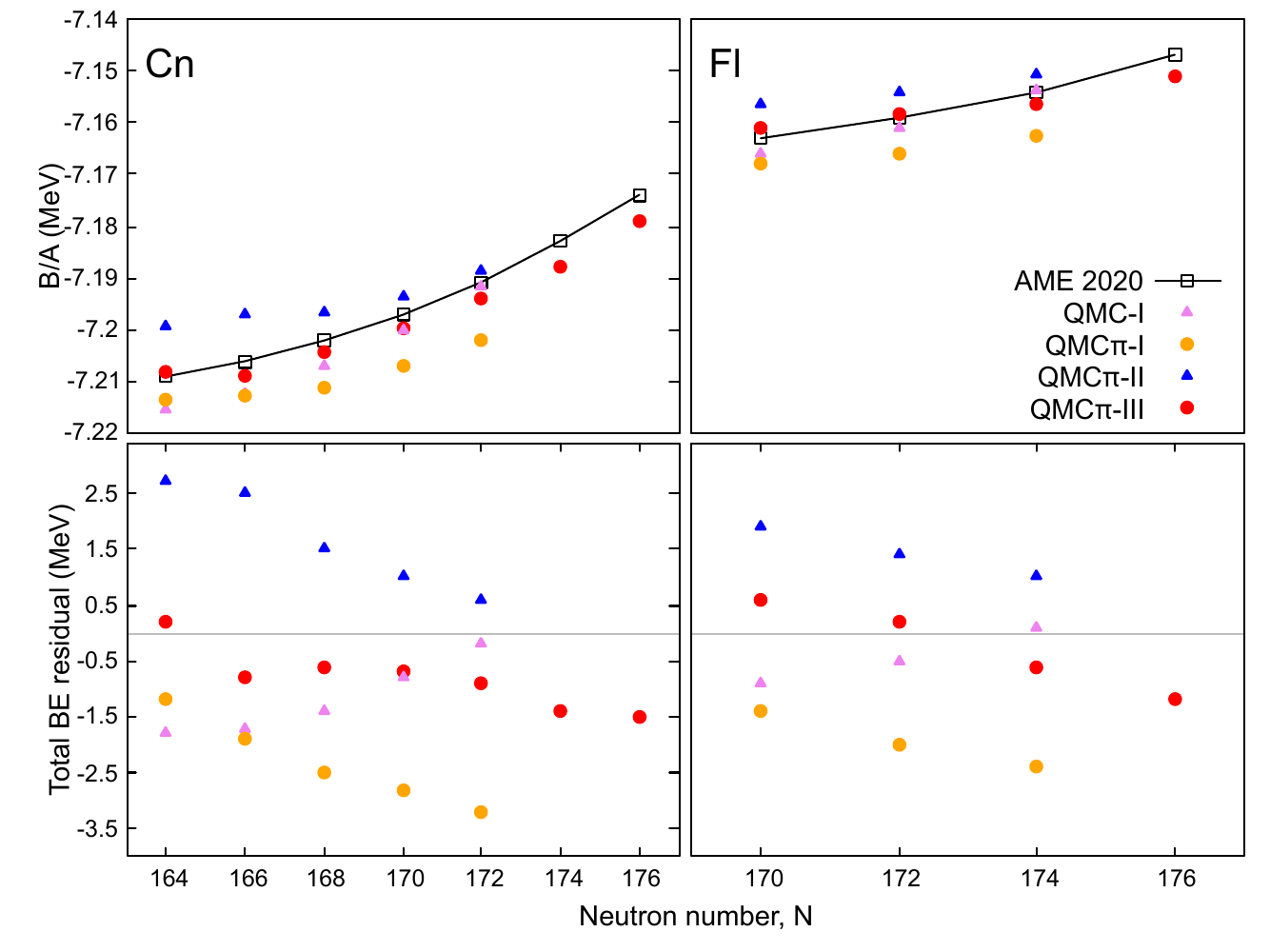}
	\end{tabular}
	\caption{$BE$ per nucleon in MeV (upper panels) and total $BE$ residuals in MeV (lower panels)  along $100\le Z \le 114$ chains computed using different QMC versions. Also shown in black empty squares are experimental data from AME 2020 \cite{AME2020}.}
	\label{BE_FmtoFl}
	\end{figure}

The nuclear binding energy (BE) is the most accessible ground-state observable. In Figure \ref{BE_FmtoFl}, the binding energy per nucleon (in MeV) and the total binding energy residuals (in MeV) are presented in the upper and lower panels, respectively, for elements with atomic numbers ranging from $Z=100$ to $Z=114$. The residuals are calculated as described in Section \ref{method} using data from AME 2020 \cite{AME2020}. For comparison, results from previous versions of the QMC model are included in the plots.

As seen in Figure \ref{BE_FmtoFl}, \qmc{III} has improved predictions for binding energies, achieving absolute total binding energy residuals of less than approximately 1.5 MeV. The relatively steep increase in binding energy per nucleon (B/A) at neutron numbers $N=152$ for $Z=100$ (Fermium), $Z=102$ (Nobelium), and $Z=104$ (Rutherfordium) isotopic chains, as well as at $N=162$ for $Z=106$ (Seaborgium), $Z=108$ (Hassium), and $Z=110$ (Darmstadtium) chains, is effectively captured by the latest QMC model. These subshell closures are further discussed in relation to alpha decay energies ($Q_\alpha$) in Section \ref{sec:Qa}.

For the isotopic chains of Livermorium ($Z=116$) and Oganesson ($Z=118$), the binding energy residuals are shown in Table \ref{ebinresid_LvOg}. It is again highlighted that the latest \qmc{III} has improved binding energy predictions, resulting in generally smaller residuals.

\begin{table}[h!]
\centering
\caption{Comparison of total $BE$ residuals for $Z \ge 116$ from different QMC versions.}
\begin{tabular}{c c c c c c c}
\hline
Element&$Z$	&	$N$	&	QMC-I&\qmc{I}&\qmc{II}&\qmc{III}\\\hline
Lv&116	&	174	&	-0.4	&	-1.2	&	2.4	&	-0.3	\\
&116	&	176	&	0.2	&	-1.6	&	2.0	&	-1.3	\\
Og&118	&	176	&	-0.6	&	-0.8	&	3.5	&	-0.8	\\
\hline
\end{tabular}
\label{ebinresid_LvOg}
\end{table}

Table \ref{BEdiff} presents the root-mean-square $BE$ residuals calculated using various nuclear model predictions and the AME 2020 \cite{AME2020} for a total of 60 known even $Z$, even $N$ SHEs. Notably, the \qmc{III} model demonstrated exceptional accuracy, yielding a root-mean-square residual of approximately 0.7 MeV, while the other models produced residuals exceeding 7 MeV.
	
\begin{table}[h!]
\centering
\caption{Comparison of root-mean-square $BE$ residuals computed using various nuclear models and the latest AME 2020 \cite{AME2020} for a total of 60 known even $Z$, even $N$ SHEs.}
\begin{tabular}{|c|c|}
\hline
Nuclear model & root-mean-square $BE$ residual \\ \hline
\qmc{III} & 0.7 MeV \\
\qmc{II} & 2.0 MeV\\
UNEDF1 & 1.4 MeV\\
DDME$\delta$ & 2.3 MeV\\
FRDM & 2.6 MeV\\
SVmin-HFBTHO & 7.1 MeV\\
\hline
\end{tabular}
\label{BEdiff}
\end{table}
	
\subsection{$Q_\alpha$ energies}
\label{sec:Qa}

In the superheavy region, most isotopes undergo alpha decay. The energy required to remove an $\alpha$ particle from the nucleus, denoted as \( Q_\alpha \), can be calculated using the energy difference between the parent isotope and the decay products, which are the daughter nucleus and the $\alpha$ particle. This relationship is expressed by the equation:

\begin{equation}
Q_\alpha(Z,N) = BE(Z,N) - BE(Z-2,N-2) - BE(2,2)\,,
\end{equation}
where $BE(2,2) = -28.296$ MeV represents the binding energy of the $\alpha$ particle (Helium-4).

\begin{figure}[!h]
		\centering
		\includegraphics[angle=0,width=0.7\textwidth]{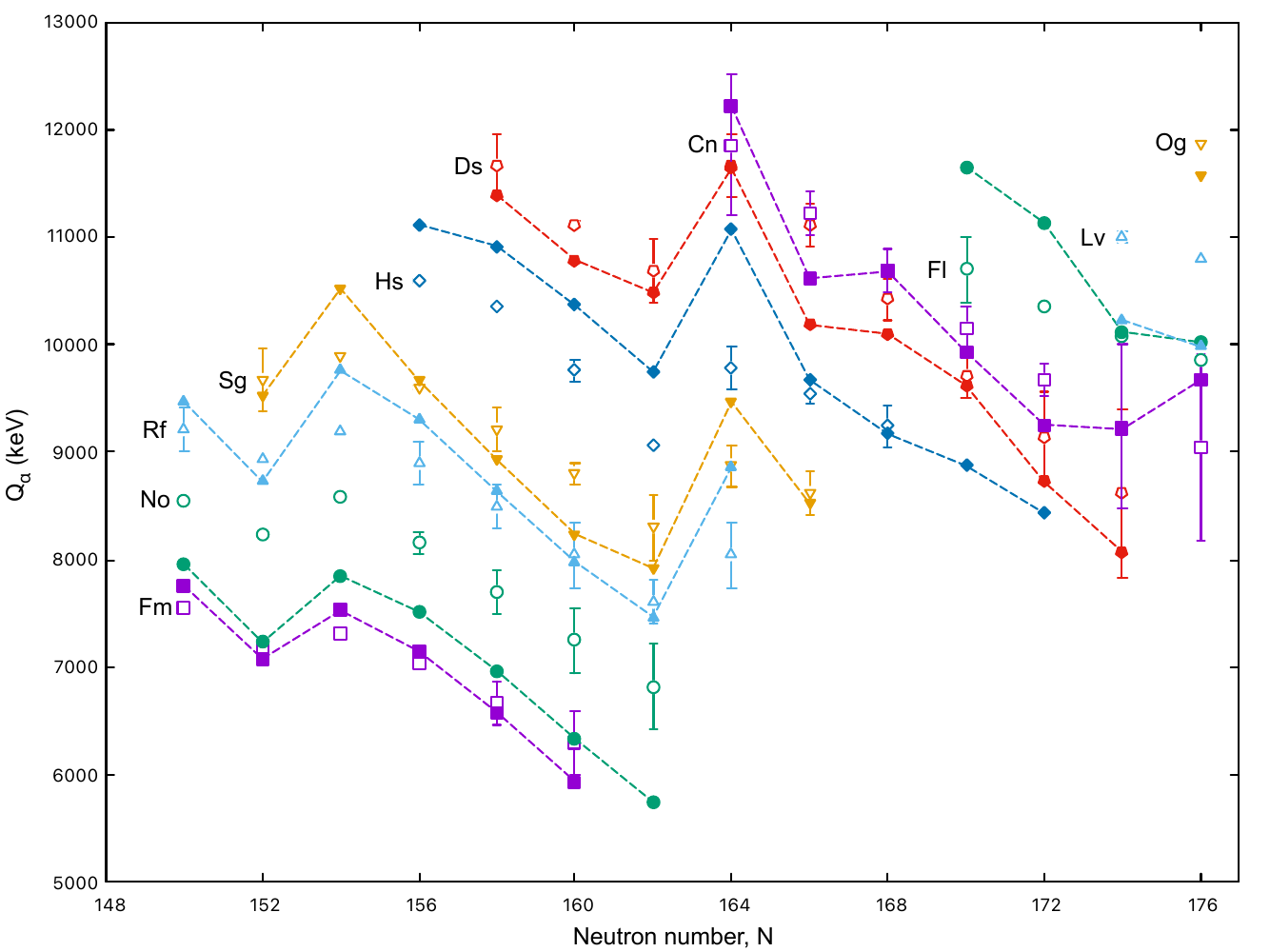}\\
		\caption{Comparison of $Q_\alpha$ energies from \qmc{III} predictions and AME 2020 \cite {AME2020} data. Results from \qmc{III} are shown as filled symbols and connected by dashed lines while experimental data and errors are shown as empty symbols with vertical error bars.}
		\label{Qa}
	\end{figure}

Figure \ref{Qa} shows the \qmc{III} predictions for $Q_\alpha$ energies, indicated by filled symbols and connected by dashed lines. In contrast, experimental data from AME 2020 \cite{AME2020} is represented by empty symbols with vertical error bars.  It is noticeable that the subshell closures at neutron numbers \( N=152 \) and \( N=162 \) are captured well by the \qmc{III} model. The closures are evident from the sudden dips in \( Q_\alpha \) energies along the isotopic chains of SHEs observed in the experimental data. At a shell closure, the \( Q_\alpha \) value is expected to be lower than that of the subsequent isotope because it takes more energy to remove an $\alpha$ particle from a nucleus immediately following a closed shell.

\subsection{Quadrupole deformations}
\begin{figure}[!ht]
	\centering
	\begin{tabular}{c c}
	\includegraphics[angle=0,width=0.485\textwidth]{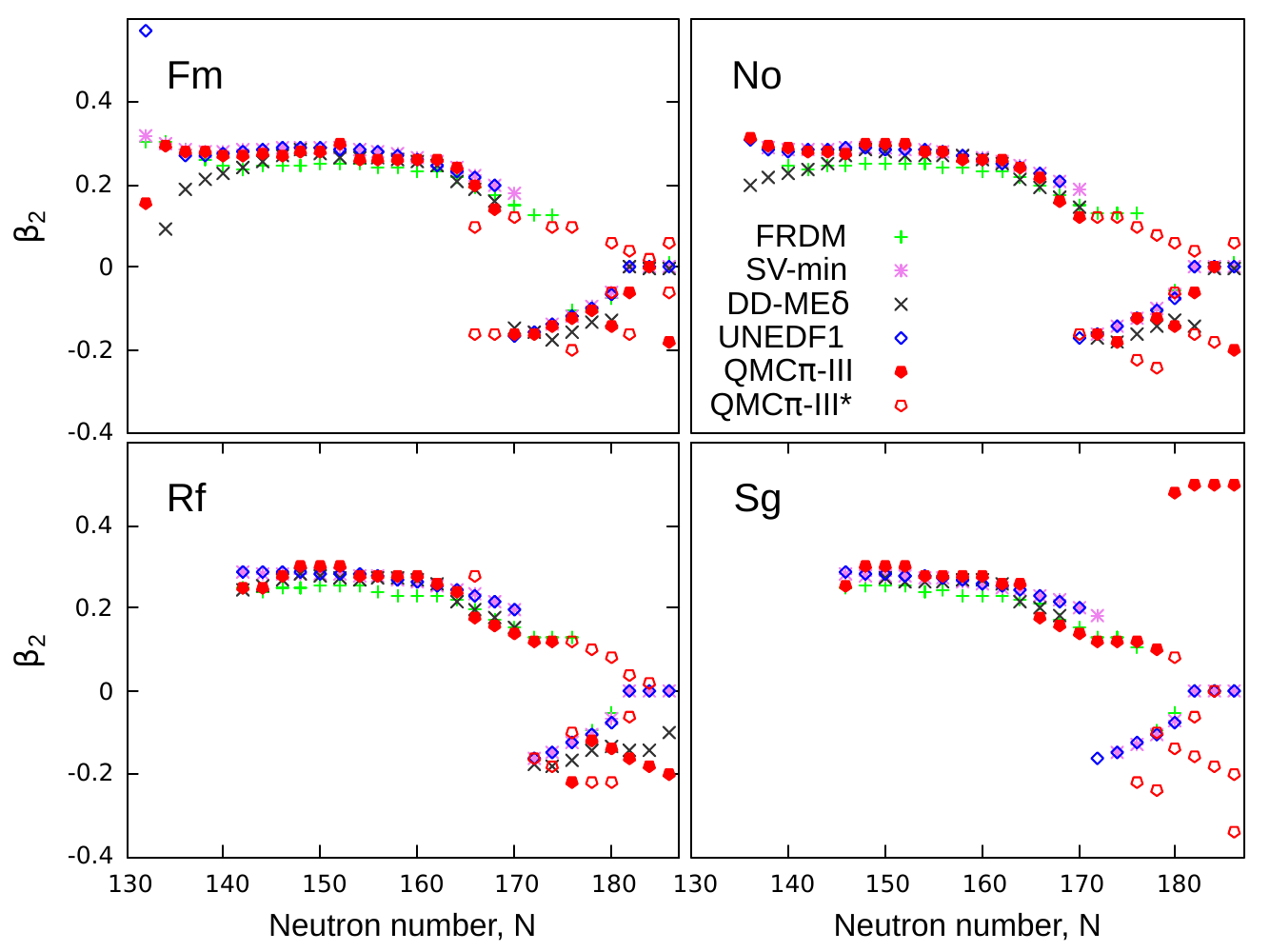} & 
	\includegraphics[width=0.485\linewidth]{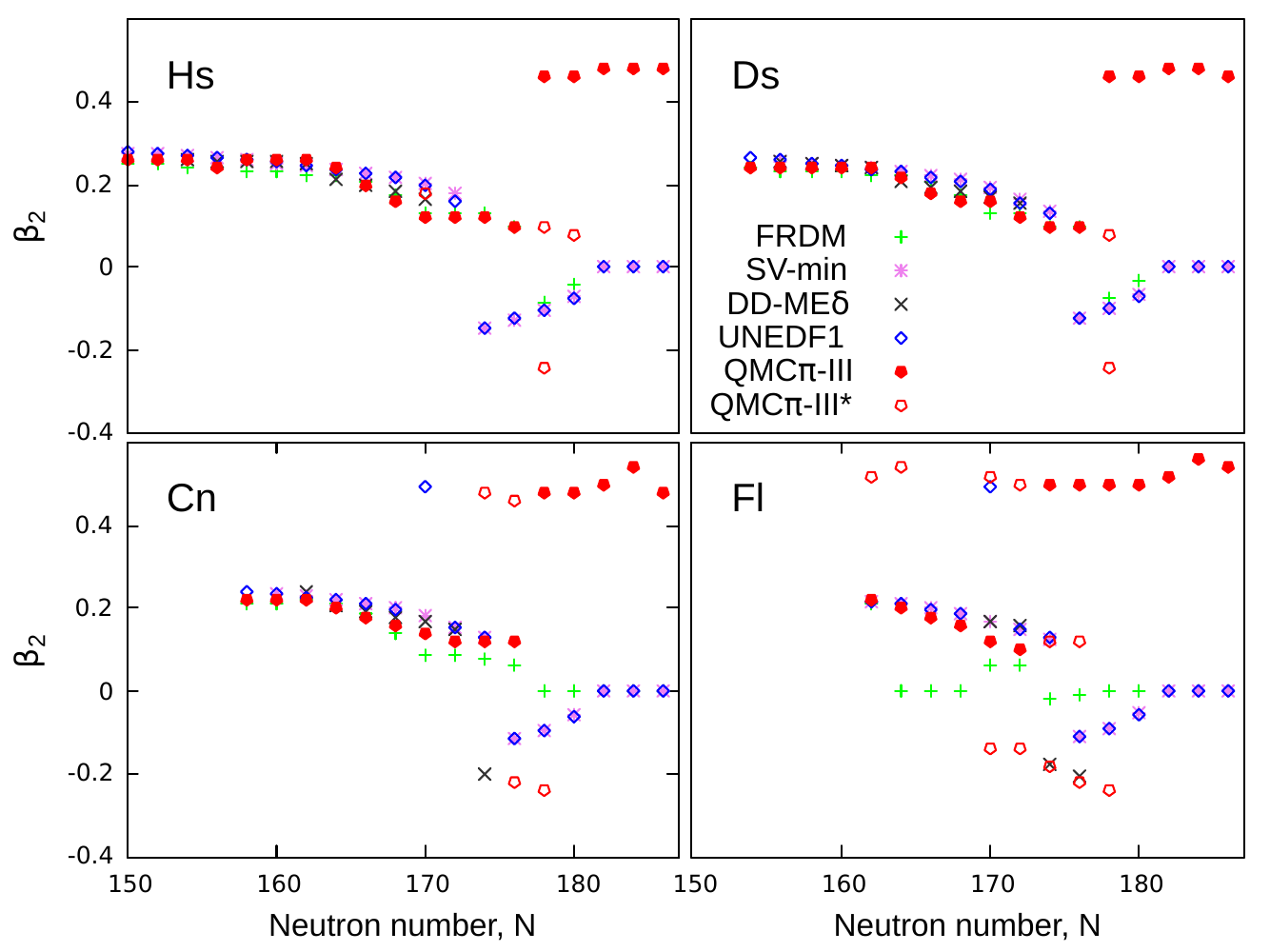}
	\end{tabular}
\caption{Comparison of deformation parameter, $\beta_2$, values along the $Z=100$ to $Z=114$ isotopic chains from several nuclear models. For \qmc{III}, the first minima are shown as filled red symbols, while the other minima, which are very close to the first, are shown as empty red symbols and labeled `\qmc{III}*'.}
	\label{deform}
\end{figure}

Currently, there is no available experimental data on the shapes of SHEs. Hence, the deformation properties predicted by the model \qmc{III} are compared with those from other nuclear models. In Figure \ref{deform}, the values of the quadrupole deformation parameter, $\beta_2$, from various models are plotted against the neutron number, $N$, for isotopic chains ranging from $Z=100$ to $Z=114$. The \qmc{III} model predicts shape coexistence, indicating that some SHEs exhibit two or more shapes corresponding to the minimum values of deformation energies. In the plots, the first minima are represented by filled red symbols, while the closely positioned second minima are shown as empty red symbols and labeled as ‘\qmc{III*}’.

The results from \qmc{III} closely align with those of other models, predicting prolate shapes up to approximately $N = 170$ for the isotopic chains of Fm, Rf, and No. Beyond that, shapes transition to oblate from $N = 172$ to $N = 180$. Additionally, \qmc{III} yields a second prolate minimum for the said isotopes within the range of $172 \le N \le 180$. At the predicted shell closure $N = 184$, \qmc{III} forecasts the coexistence of spherical and oblate shapes, while most other models predict only spherical shapes. For the Sg chain, the primary minimum suggests a transition from prolate to spherical shapes up to around $N = 180$, with even more prolate shapes appearing further along the chain.

For the Hs, Ds, Cn, and Fl chains, \qmc{III} predominantly predicts prolate shapes, with the development of shape coexistence as the atomic number $Z$ increases. However, at the anticipated $N = 184$ closure, \qmc{III} yields predominantly prolate shapes, while other models continue to predict spherical shapes for the $N=184$ isotones.

\section{Summary and concluding remarks}
Much is anticipated in the superheavy region of the nuclear chart as scientists explore what lies ahead. Since there is limited available data, theoretical predictions rely on nuclear models that are continually refined in hopes of better understanding nucleon-to-nucleon interactions, especially in the presence of very large numbers of protons and neutrons.

The QMC model has demonstrated impressive results in this area of the nuclear chart, particularly in its predictions for energy observables. The model yields residuals of less than 1.5 MeV for total binding energies and provides accurate estimates for $Q_\alpha$ energies, thereby capturing the expected subshell closures. Additionally, the QMC model is capable of delivering comparable results for nuclear shapes and deformations despite significantly fewer model parameters than other nuclear models. It is, therefore, imperative to conduct further theoretical exploration using the QMC model in regions beyond the current landscape to drive significant insights and advancements.

\section*{Acknowledgements}
Pierre A. M. Guichon and Jirina R. Stone would like to acknowledge the hospitality of the CSSM at the University of Adelaide. This work was supported by the Australian Research Council through Discovery Project DP230101791 (AWT).

\end{document}